\documentclass[%
 reprint,
 superscriptaddress,
 amsmath,amssymb,
 aps,
]{revtex4-2}

\usepackage{stmaryrd}
\usepackage{graphicx}
\usepackage{dcolumn}
\usepackage{bm}
\usepackage{physics} 
\usepackage{braket}
\usepackage{hyperref}
\hypersetup{
  colorlinks   = true,    
  urlcolor     = blue,    
  linkcolor    = blue,    
  citecolor    = blue      
}
\usepackage[capitalize]{cleveref}
\usepackage{accents}
\usepackage{enumerate}
\usepackage[T1]{fontenc}
\usepackage{tabularx}

\DeclareDocumentCommand\ceil{ s m }
{
    \lceil #2 \rceil
}
\DeclareDocumentCommand\floor{ s m }
{
    \lfloor #2 \rfloor
}

\usepackage{xcolor}
\definecolor{myyellow}{HTML}{f1a226}
\definecolor{mygreen}{HTML}{298c8c}
\usepackage{graphicx}
\usepackage{placeins}

\begin{document}

\preprint{APS/123-QED}

\title{Crosstalk in Multi-Qubit Fluxonium Architectures with Transmon Couplers}

\author{Martijn F. S. Zwanenburg}
\email[Contact author: m.f.s.zwanenburg@tudelft.nl]{}
\affiliation{QuTech and Kavli Institute of Nanoscience, Delft University of Technology, 2628 CJ, Delft, The Netherlands}
\author{Christian Kraglund Andersen}
\email[Contact author: c.k.andersen@tudelft.nl]{}
\affiliation{QuTech and Kavli Institute of Nanoscience, Delft University of Technology, 2628 CJ, Delft, The Netherlands}

\date{\today}

\begin{abstract}
In recent years, several architectures have been proposed for implementing two-qubit operations on fluxonium superconducting qubits. A particularly promising approach, which was demonstrated experimentally by Refs. \cite{mit_cz, sidd_eugene}, employs a transmon superconducting qubit as a tunable coupler between the fluxonium qubits. These experiments have shown that the transmon coupler enables fast, high-fidelity two-qubit operations while suppressing unwanted ZZ crosstalk between the fluxonium qubits. In this work, we numerically study the scalability of this architecture. We find that, when trivially scaling this architecture, crosstalk from spectator qubits limits the gate fidelity to below 90\%. We show that these spectator errors can be reduced to below $10^{-4}$ by reducing the coupling strength and by dynamically tuning transmons that are not used for a two-qubit operation to an off position. We further investigate the resilience of the operation to direct capacitive coupling between the transmon couplers and to microwave crosstalk. 
\end{abstract}

\maketitle


\section{Introduction}
\label{sec:introduction}
Recent progress with reset \cite{fluxonium_initialization}, readout \cite{fpa_experimental, nesterov_readout}, single-qubit operations \cite{circular_polarized_drive, sqgbrwa} and two-qubit operations \cite{microwave_activated_cz, fluxonium_tunable_coupler, mit_cz, fluxonium_cross_resonance, sidd_eugene, stable_cnot, fx_double_tmon} on fluxonium superconducting qubits \cite{fluxonium} have made this qubit a promising building block for larger scale quantum devices. One key area that has not yet received sufficient attention is the scaling of fluxonium qubits beyond two-qubit architectures, and in this work we are specifically interested in scaling the fluxonium-transmon-fluxonium (FTF) architecture realized experimentally by Refs. \cite{mit_cz, sidd_eugene}. In larger scale systems, the frequencies of one FTF pair may depend on the state of other fluxonium qubits (spectators), which gives rise to crosstalk errors \cite{spectators1, spectators2, spectators3}. In this work, we numerically investigate whether larger scale systems, such as building blocks for 2D surface codes \cite{surfacecode_eth, surfacecode_delft, surfacecode_google2, surfacecode_google} in which a two-qubit operation may be influenced by six nearest-neighbor fluxoniums, can effectively suppress these crosstalk errors. Previous studies of crosstalk errors on larger-scale superconducting systems were focused on transmon qubits \cite{dmrgx_google} or did not explicitly analyze the two-qubit gate errors arising in such systems \cite{dmrgx_hartmann}. 

\begin{figure}[t]
\centering
\includegraphics[width=\linewidth]{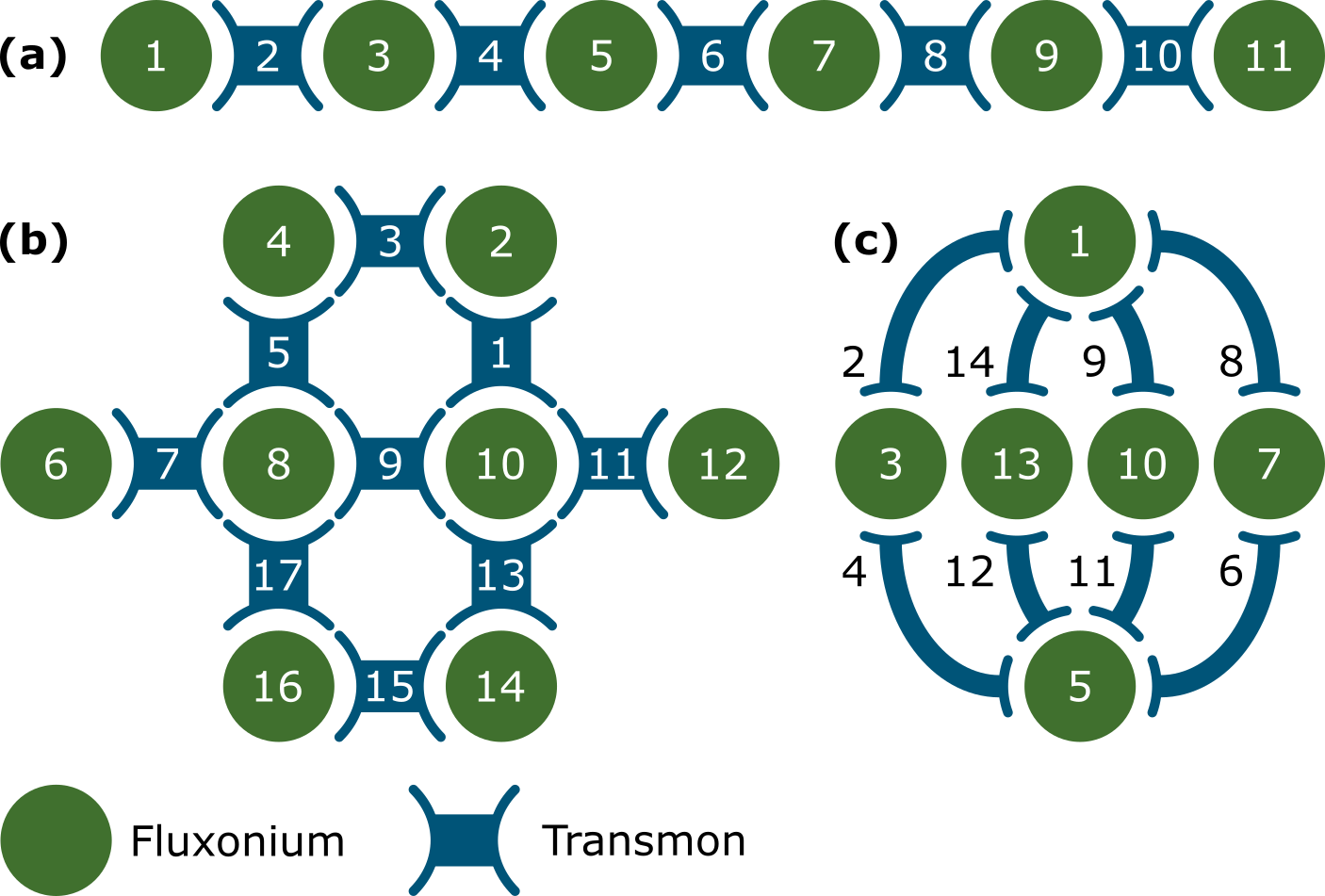}
\caption{\label{fig:grid1d} Schematic drawing of the systems studied in this work. The numbers indicate the site numbering used throughout the main text. (a) 1D system, (b) 2D grid system and \mbox{(c) 2D} system for the $\llbracket 4,2,2 \rrbracket$ error correcting code with two auxiliary qubits.}
\end{figure}

In the FTF architecture, a transmon qubit \cite{transmon} is used as a tunable coupling element between two fluxonium qubits. The transmon suppresses unwanted ZZ coupling between the fluxoniums, while at the same time enabling the implementation of a conditional phase gate (CZ). The ZZ suppression is realized by balancing the direct coupling between the fluxoniums and the indirect coupling mediated by the transmon \cite{mit_cz, sidd_eugene}. The two-qubit operation is applied by conditionally driving a $2\pi$ rotation between a computational and non-computational state. During this operation, the state picks-up a geometrical phase of $\pi$, as required for the CZ gate. Given that the two-qubit gate relies directly on the resonant driving of a specific transition, it is clear that frequency shifts induced by nearby spectator qubits may indeed cause gate errors. In this setting, we are interested in the spectator-qubit-state-dependent transition frequencies of the transmon coupler. We study the gate errors that arise from spectator effects in the 1D and 2D systems shown in \mbox{Fig. \ref{fig:grid1d}} with system parameters inspired by the experimental implementations by Refs. \cite{mit_cz, sidd_eugene}. Our results show that large crosstalk errors arise in these systems that limit the two-qubit gate fidelity to below 90\%. On the other hand, by dynamically tuning transmon couplers that are not actively used for a two-qubit operation down in frequency, we show that these crosstalk errors can be suppressed to below $10^{-4}$. 

In the remainder of this work, we first summarize the most relevant properties of the FTF system in Sec. \ref{sec:ftf}. Next, in Sec. \ref{sec:spec_errors} we study the crosstalk errors in the systems shown in \mbox{Fig. \ref{fig:grid1d}}. Finally, in Sec. \ref{sec:model_imperfections} we analyze the effect of direct capacitive coupling between the transmon couplers and microwave crosstalk on the two-qubit gate operation. 

\section{FTF Architecture}
\label{sec:ftf}
\begin{figure}[t]
\centering
\includegraphics[width=\linewidth]{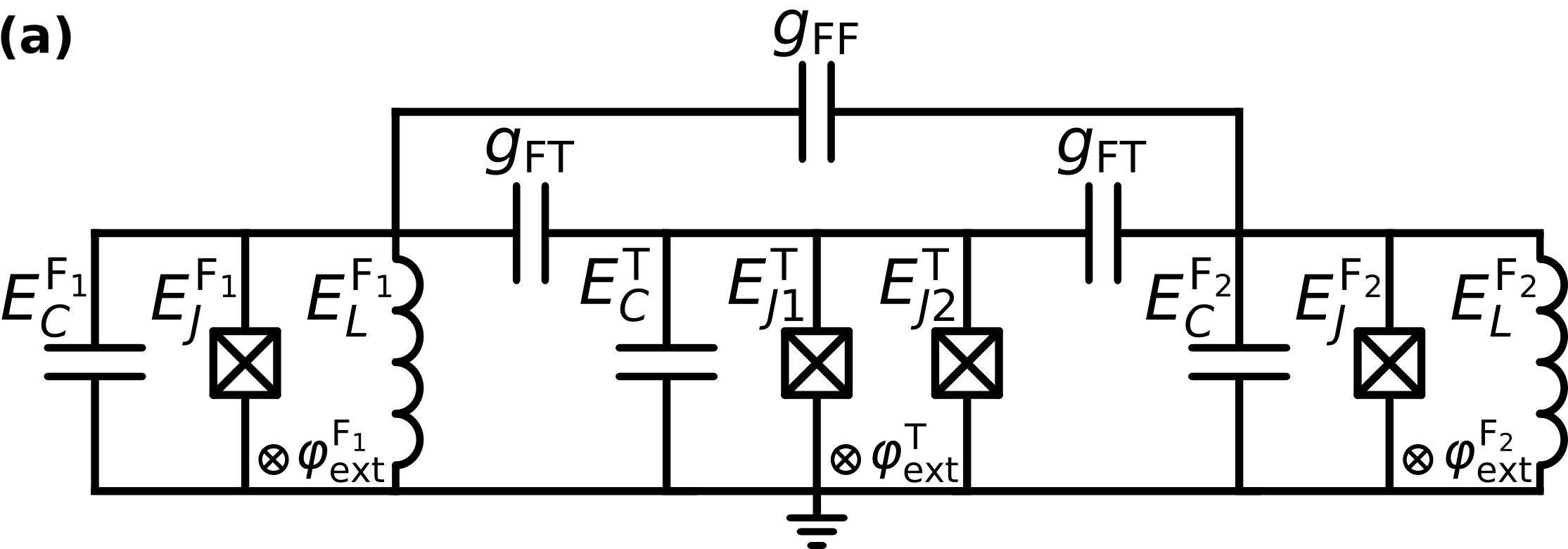}\\
\vspace{1em}
\includegraphics[width=\linewidth]{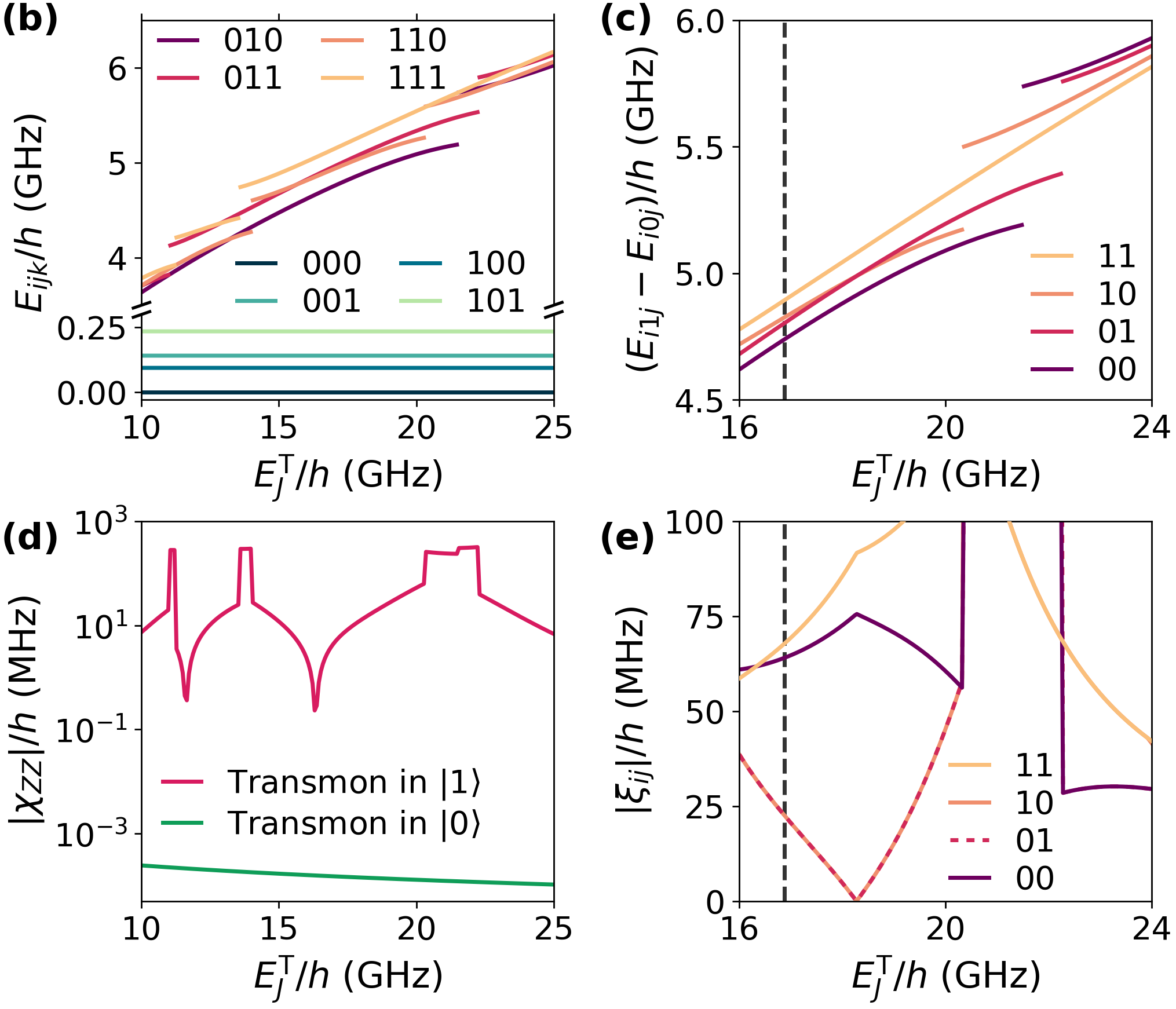}
\caption{\label{fig:ftf_schematic} (a) Schematic circuit diagram of the three node FTF system. (b) Fluxonium and transmon energy levels as a function of $E_{J}^\text{T}$. The green-shaded and red-shaded lines correspond to states with the transmon in its ground state and first excited state respectively. (c) Transition energies of the transmon's qubit level as a function of $E_{J}^\text{T}$. (d) Residual ZZ coupling between the fluxonium qubits with the transmon in the ground or excited state. (e) Minimum frequency separation of the transmon's qubit transition. The vertical dashed line in panels (c) and (e) indicate the $E_{J}^\text{T}$ experimentally realized in Ref. \cite{sidd_eugene}.}
\end{figure}

We consider the case where the fluxonium qubits are capacitively coupled to the transmon coupler, see \mbox{Fig. \hyperref[fig:ftf_schematic]{\ref*{fig:ftf_schematic}(a)}}. The individual fluxonium qubits are described by their Hamiltonian

\begin{equation}
\label{eq:HFX}
    \hat{H}_{0}^{\text{F}_i} = \; 4E_{C}^{\text{F}_i}\hat{n}_{\text{F}_i}^2 + \frac{1}{2}E_{L}^{\text{F}_i} \hat{\phi}_{\text{F}_i}^2 - E_{J}^{\text{F}_i} \cos(\hat{\phi}_{\text{F}_i} - 2\pi\varphi_{\text{ext}}^{\text{F}_i}),
\end{equation}

\noindent where $E_{C}^{\text{F}_i}$, $E_{L}^{\text{F}_i}$ and $E_{J}^{\text{F}_i}$ are the capacitive, inductive and Josephson energy of fluxonium $i$ respectively, and where $\hat{n}_{\text{F}_i}$ and $\hat{\phi}_{\text{F}_i}$ are the charge and phase operators of the fluxonium. Within this work, we only consider the case where the fluxonium qubits are flux-biased to their sweet spot, i.e., the reduced external flux is fixed at $\varphi_{\text{ext}}^{\text{F}_i} = 0.5$. The transmon is similarly described in terms of its charge operator $\hat{n}_\text{T}$ and phase operator $\hat{\phi}_\text{T}$ by the Hamiltonian

\begin{equation}
\label{eq:HT}
    \hat{H}_{0}^{\text{T}} = \; 4E_{C}^{\text{T}}\hat{n}_\text{T} - E_{J}^{\text{T}}(\varphi_\text{ext}^\text{T}) \cos(\hat{\phi}_\text{T}),
\end{equation}

\noindent where $E_{C}^{\text{T}}$ is the fixed charging energy and where

\begin{equation}
\begin{split}
    E_{J}^{\text{T}}(\varphi_\text{ext}^\text{T})= (E_{J1}^{\text{T}} + E_{J2}^{\text{T}})\sqrt{\cos(\pi\varphi_\text{ext}^\text{T})^2 + d^2\sin(\pi\varphi_\text{ext}^\text{T})^2},
\end{split}
\end{equation}

\noindent is the effective Josephson energy of the transmon's SQUID loop that is threaded by a reduced external flux $\varphi_\text{ext}^\text{T}$ \cite{transmon}. Here, $d=\abs{E_{J1}^\text{T} - E_{J2}^\text{T}}/(E_{J1}^\text{T} + E_{J2}^\text{T})$ denotes the SQUID asymmetry of the transmon. Given the capacitive coupling between the systems, the coupling Hamiltonians are given by

\begin{equation}
\label{eq:HC}
\begin{split}
    \hat{H}_\text{FT} =& \; \hbar g_\text{FT}\Big(\hat{n}_{\text{F}_1}\hat{n}_\text{T} + \hat{n}_\text{T}\hat{n}_{\text{F}_2}\Big), \\
    \hat{H}_\text{FF} =& \; \hbar g_\text{FF}\hat{n}_{\text{F}_1}\hat{n}_{\text{F}_2},
\end{split}
\end{equation}

\noindent where $g_\text{FT}$ is the coupling strength between the transmon and each fluxonium, and $g_\text{FF}$ is the coupling strength between the fluxoniums. The total Hamiltonian is given by

\begin{equation}
\label{eq:three_body_ham}
    \hat{H} = \sum_{i \in \{\text{F}_1,\text{T},\text{F}_2\}} \hat{H}_{0}^i + \hat{H}_\text{FT} + \hat{H}_\text{FF}
\end{equation}

\noindent In this work, we will denote eigenstates of this three node Hilbert space as $\ket{\text{F}_1\text{TF}_2}$ with corresponding eigenenergies $E_{\text{F}_1\text{TF}_2}$. 

When using the system parameters realized by \mbox{Ref. \cite{sidd_eugene}}, the spectrum of the low-energy subspace readily reveals the four computational levels while we see a higher energy subspace consisting of four non-computational states when the transmon is excited, see Fig. \hyperref[fig:ftf_schematic]{\ref*{fig:ftf_schematic}(b)}. As $E_{J}^\text{T}$ increases, the resulting transmon transition frequencies become resonant with the $\ket{1}\leftrightarrow\ket{2}$ and the $\ket{0}\leftrightarrow\ket{3}$ transitions of the fluxoniums, which we see in Fig. \hyperref[fig:ftf_schematic]{\ref*{fig:ftf_schematic}(b)} as avoided crossings in the non-computational states. Avoided crossing with the $\ket{0}\leftrightarrow\ket{2}$ and $\ket{1}\leftrightarrow\ket{3}$ transitions of the fluxoniums do not appear, as these transitions are parity forbidden when the fluxonium is parked at its sweet spot. Notice that, for simplicity, we have omitted the explicit external flux dependence of $E_{J}^\text{T}$. In particular, we see in Fig. \hyperref[fig:ftf_schematic]{\ref*{fig:ftf_schematic}(c)} that the transmon transition frequencies are different for each of the four possible two-qubit states of the fluxonium, which enables a microwave-activated conditional phase gate between the fluxonium qubits. The dashed line in \mbox{Fig. \hyperref[fig:ftf_schematic]{\ref*{fig:ftf_schematic}(c)}} marks the operating point of Ref. \cite{sidd_eugene}. The residual ZZ coupling between the fluxonium qubits, defined as $\chi_{ZZ} = E_{1i1} - E_{0i1} - E_{1i0} + E_{0i0}$, remains below 1 kHz for a wide range of transmon frequencies provided that the transmon is in its ground state, see \mbox{Fig. \hyperref[fig:ftf_schematic]{\ref*{fig:ftf_schematic}(d)}}. When the transmon is in $\ket{1}$ the ZZ coupling is significantly larger as a result of the coupling between the non-computational states.

It is important to note that, in principle, different choices exist with regards to the exact transition that is driven for the conditional $2\pi$ rotation that implements the CZ operation. For instance, in Refs. \cite{mit_cz,sidd_eugene} the gate was implemented by driving the $\ket{101} \leftrightarrow \ket{111}$ transition, but the gate can also be implemented by driving the $\ket{101} \leftrightarrow \ket{201}$ or $\ket{101} \leftrightarrow \ket{102}$ transitions as also done in Ref. \cite{mit_cz}. In this work, we will focus on implementing the CZ by driving a transmon transition, i.e. $\ket{i0j} \leftrightarrow \ket{i1j}$. As also discussed in Ref. \cite{sidd_eugene}, the speed of the operation is limited by residual, undesired driving of other transitions that are close in energy to the targeted transition, which results in leakage. When driving a transmon transition $\ket{i0j} \leftrightarrow \ket{i1j}$, there are numerous leakage transitions that can be driven off-resonantly, for instance the other three transmon transitions shown in Fig. \hyperref[fig:ftf_schematic]{\ref*{fig:ftf_schematic}(c)}. Therefore, a key parameter that influences the speed of the operation is the minimum frequency separation $\xi_{ij}$

\begin{equation}
    \label{eq:conditional_frequency}
\begin{split}
    \xi_{ij} = \min_{kl \neq ij } \big\vert & \big(E_{k1l}-E_{k0l} \big) - \big(E_{i1j} - E_{i0j} \big) \big\vert,
\end{split}
\end{equation}

\noindent which specifies the minimum frequency difference of the targeted transmon transition to the other transmon transitions. A larger minimum frequency separation indicates smaller leakage errors and hence a potentially faster gate. In \mbox{Fig. \hyperref[fig:ftf_schematic]{\ref*{fig:ftf_schematic}(e)}}, we see that the largest minimum frequency separation for the $\ket{101} \leftrightarrow \ket{111}$ transition is achieved when tuning the transmon as close as possible to the avoided crossings. However, $\xi_{ij}$ does not provide a complete picture of the total leakage in the operation. For instance, tuning too close to the avoided crossings would result in leakage to the third excited state of the fluxonium through the transitions $\ket{i00} \leftrightarrow \ket{i03}$ and $\ket{00j} \leftrightarrow \ket{30j}$. Furthermore, the extend to which a leakage transition is driven not only depends on the frequency difference, but also on how strongly it couples to the drive operator. 

In this two-qubit gate scheme, the residual ZZ interaction remains suppressed at the operating point. Therefore, in the experimental implementations of Refs. \cite{mit_cz, sidd_eugene} it was not required to dynamically flux-tune the transmon from an \textit{off} to an \textit{on} point, as often required in tunable-coupler schemes \cite{tmon_tunable_coupler,mit_tft}. A potential downside is that the interaction between the non-computational states is always on, which may lead to crosstalk errors.

\section{Spectator Errors}
\label{sec:spec_errors}
To determine the spectator errors that arise when scaling-up the systems in Refs. \cite{mit_cz, sidd_eugene}, we first study the 1D architecture shown in \mbox{Fig. \hyperref[fig:grid1d]{\ref*{fig:grid1d}(a)}} consisting of six fluxoniums and five transmon couplers. We consider the application of a two-qubit operation on the center FTF pair, such that there are two spectator fluxoniums to the left and right. The parameters of the center FTF pair are set to the parameters realized in Refs. \cite{mit_cz, sidd_eugene}. The parameters for the other fluxoniums and transmons are chosen in a ladder-like fashion, such that each pair of fluxonium qubits has a similar detuning. Note that this approach does not scale indefinitely since there may be a practical upper or lower bound to the qubit frequencies. The precise system parameters used for the simulations are detailed in Appendix \ref{app:system_params}. We highlight that the key difference between the parameter regimes used in \mbox{Refs. \cite{mit_cz, sidd_eugene}} are the coupling strengths, which are $g_\text{FT}/2\pi=550$ MHz and $g_\text{FT}/2\pi=236$ MHz, respectively. 

To analyze the spectator errors, we first need to compute the eigenvalues and eigenvectors of the system. We compute the spectrum for each individual fluxonium and transmon, and truncate their Hilbert spaces to the lowest six and three levels, respectively. For the 1D system, the dimension of the resulting combined truncated Hilbert space is approximately $11\cdot 10^6$, which is roughly equivalent to the Hilbert space dimension of 23 qubits. It is therefore highly challenging to compute the spectrum with exact diagonalization, and instead we employ a tensor network algorithm in which the eigenstates are represented as a matrix-product state (MPS) \cite{schollwock, tensor_networks}. Specifically, we make use of the density matrix renormalization group for excited states (DMRG-X) algorithm \cite{DMRGX}, which is a variational algorithm in which each site of the MPS is updated iteratively until it converges to an eigenstate of the system. The full implementation of this algorithm is detailed in \mbox{Appendix \ref{app:dmrgx}}.

In the FTF architecture, spectator errors arise through the dependence of the frequency and drive matrix element of the conditionally driven transition and of the dominant leakage channels on the state of spectator qubits. For the 1D architectures inspired by the parameters used in Refs. \cite{mit_cz, sidd_eugene}, we find that the energy of the $\ket{101} \leftrightarrow \ket{111}$ transition changes by \mbox{8.6 MHz} and \mbox{5.7 MHz}, respectively, when the four spectator fluxoniums are flipped from the $\ket{0}$ state to the $\ket{1}$ state. To quantify the resulting crosstalk from spectator qubits in terms of gate errors, we numerically solve the time evolution of the two-qubit gate. Due to the size of the system and the strong interactions, it is infeasible to time evolve the full system in a tensor network formalism, and instead we perform a truncation to the center FTF pair where the operation is performed. First, we transform the three-node Hamiltonian in Eq. \eqref{eq:three_body_ham} to the eigenframe, and add a drive Hamiltonian to obtain

\begin{equation}
\label{eq:three_node_dh}
\begin{split}
    \hat{\tilde{H}}/\hbar &= \sum_i \omega_i \ket{i}\bra{i} + \hat{\tilde{H}}_D/\hbar, \\
    \hat{\tilde{H}}_D/\hbar &= \Omega\mathcal{E}_I(t)\cos(\omega_dt)\hat{\tilde{D}},
\end{split}
\end{equation}

\noindent where $\omega_i$ specifies the energy of each level, $\Omega$ the drive strength, $\mathcal{E}_I(t)$ the pulse envelope, $\omega_d$ the drive frequency and $\hat{\tilde{D}}$ the drive operator. Here, we drive the operation through the transmon's charge operator and neglect microwave crosstalk, such that $\hat{\tilde{D}} = \hat{\tilde{n}}_\text{T}$. The $\tilde{\cdot}$ denotes that the operator has been transformed to the eigenbasis. Using the DRMG-X algorithm, we compute the eigenenergies $\omega_i$ and we can further calculate the matrix elements of the drive operator $\hat{\tilde{D}}$ for the full 1D system. As we expect the spectator effects to be dominated by changes in the transition frequency and drive matrix elements of the transmon's qubit transition, we only compute the energies for the states $\ket{\psi(s_l,i,k,j,s_r)}=\ket{s_l0s_l0ikj0s_r0s_r}$ and matrix elements $\bra{\psi(s_l,i,1,j,s_r)}\hat{\tilde{D}}\ket{\psi(s_l,i,0,j,s_r)}$ with $i,j,k \in \{0,1\}$ for a specific configuration of the left ($s_l$) and right ($s_r$) spectator fluxoniums. The notation of the kets $\ket{\cdot}$ follows the site ordering of the 1D chain shown in \mbox{\mbox{Fig. \hyperref[fig:grid1d]{\ref*{fig:grid1d}(a)}}}. Notice that, for simplicity, we assume that the left two and right two spectator fluxoniums are in the same state. Thus, using these DMRG-X results, we can readily reconstruct the Hamiltonian in Eq. \eqref{eq:three_node_dh} to obtain the effective three node Hamiltonian for the center FTF pair for a certain spectator configuration. The resulting system is solved numerically to obtain the time evolution $U$. We highlight that this approach neglects any leakage outside the center FTF pair and neglects any spectator-dependent leakage to the second and third excited states of the transmon and fluxoniums. For systems where the eigenstates delocalize strongly, this approach may therefore only provide a lower bound on the actual crosstalk errors. 

\begin{figure}[t]
\centering
\includegraphics[width=\linewidth]{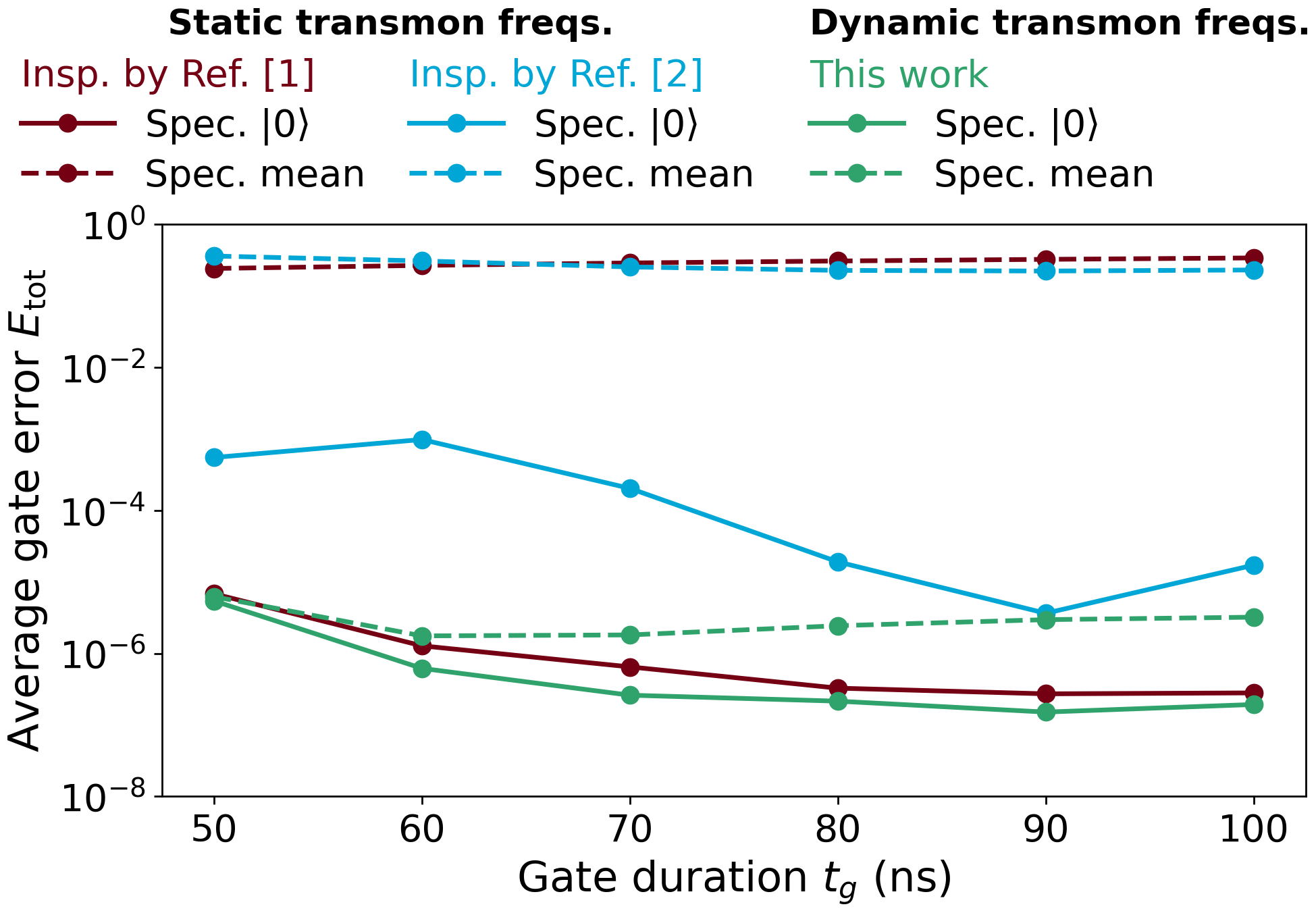}
\caption{\label{fig:spec1d} Spectator errors in a 1D chain consisting of six fluxoniums and five transmon couplers as a function of the gate duration $t_g$. For each system, the two-qubit operation is calibrated with all spectators in $\ket{0}$ (solid lines), and then the error is averaged over the four spectator states with $s_l,s_r \in \{0,1\}$ (dashed lines). In red and blue, we plot the errors for the parameters inspired (Insp.) by Refs. \cite{mit_cz} and \cite{sidd_eugene} respectively, and in green we plot the results for the parameters proposed in this work. For the parameters inspired by Refs. \cite{mit_cz, sidd_eugene} all transmons are statically positioned at their on point, while for the parameters proposed in this work the frequencies of inactive transmons are dynamically tuned to an off position.}
\end{figure}

We calculate the gate error for the 1D system with the system parameters inspired by Refs. \cite{mit_cz, sidd_eugene} as a function of the gate duration $t_g$ to elucidate the trade-off between fast driving and potential errors arising from nearby off-resonant transitions, see Fig. \ref{fig:spec1d}. Specifically, we calculate the total gate error as $E_\text{tot}=E_{\text{phase}} + E_{\text{leakage}}$, where we have defined the coherent phase error $E_{\text{phase}}$ and the average leakage error $E_{\text{leakage}}$ as

\begin{equation}
\label{eq:error_definition}
\begin{split}
    & E_{\text{phase}} = \frac{3}{10}\left(1 - \cos\left(\tilde{\theta}_c\right)\right), \\
    & E_{\text{leakage}} = 1 - \frac{1}{4}\sum_{\psi=000,001,100,101} \vert\bra{\psi}U\ket{\psi}\vert^2.
\end{split}
\end{equation}

\noindent Here, $\tilde{\theta}_c=\theta_c-\pi$ is the conditional phase error, with $\theta_c$ the conditional phase defined as

\begin{equation}
\label{eq:conditional_phase}
\begin{split}
    \theta_c = & \text{Arg}\left(\bra{101}U\ket{101}\right) - \text{Arg}\left(\bra{001}U\ket{001}\right) \\
    & - \text{Arg}\left(\bra{100}U\ket{100}\right) + \text{Arg}\left(\bra{000}U\ket{000}\right),
\end{split}
\end{equation}

\noindent where $\text{Arg}(\cdot)$ denotes the argument. We implement the CZ gate by driving the $\ket{101}\leftrightarrow\ket{111}$ transition with a cosine pulse defined as

\begin{equation}
\label{eq:consine_pe}
    \mathcal{E}_I(t) = \frac{1}{2}\left(1 - \cos\left(\frac{2\pi}{t_g}t\right)\right).
\end{equation}

\noindent Importantly, we numerically calibrate the drive frequency $\omega_d$ and drive strength $\Omega$ to minimize the gate error when all spectator qubits are in the state $\ket{0}$. More details regarding the drive pulse and calibration can be found in Appendix \ref{app:calibration}. Next, we calculate the error averaged over four different spectator states $s_l,s_r \in \{0,1\}$. We observe that, while the error with the spectators in $\ket{0}$ is low, the error averaged over the spectator states is above 0.1 for all gate durations. The individual contributions of the phase and leakage errors to the total error are plotted in \mbox{Appendix \ref{app:calibration}}, from which we conclude that, while the phase errors are the limiting error source, both leakage and phase errors are above or around 1\%.

The results in Fig. \ref{fig:spec1d} show that trivially scaling the parameters of Refs. \cite{mit_cz, sidd_eugene} is not feasible due to large spectator errors. We expect these errors to increase even further when considering a 2D architecture, where the two-qubit operation is subject to six nearest-neighbor spectator qubits instead of the two nearest-neighbor spectator qubits in 1D. Note that we have only considered a simple pulse shape here and we would expect that pulse shaping techniques such as derivative removal by adiabatic gate (DRAG) and related techniques \cite{drag1,drag2,hd-drag} could reduce errors for the calibrated gate with all spectators in $\ket{0}$ \cite{sidd_eugene}. Importantly, however, we do not expect such pulse shaping techniques to reduce spectator errors since they are not designed to be robust against changes in the primary transition frequency.

Next, our goal is to suppress the spectator errors. We readily understand that the large spectator errors arise from the transmon-like modes being delocalized over multiple couplers. Thus, we will achieve the suppression by reducing the coupling $g_\text{FT}$ and by increasing the frequency spacing between the fluxonium and transmon transitions. The full parameter regime, which we found to reduce the spectator effects, is detailed in \mbox{Appendix \ref{app:system_params}}. Most notably, we reduce the coupling to $g_\text{FT}/2\pi=200$ MHz and we dynamically tune the frequency of \textit{inactive} transmons, i.e., transmons that are not actively used to perform a two-qubit gate, away from the $\ket{0} \leftrightarrow \ket{3}$ transition of the fluxoniums to suppress the residual coupling between these levels. Specifically, we tune these transmons down in frequency to $\sim$300 MHz above the fluxoniums' $\ket{1} \leftrightarrow \ket{2}$ transition. Experimentally, this can be realized by flux-pulsing the transmon. For the 1D chain, only the transmons that sit directly next to the center FTF pair are tuned to this off point, such that simultaneous two-qubit operations can be performed on the middle and two outer FTF pairs. We highlight that, at their off position, the transmon couplers still suppress the ZZ interaction between the fluxoniums to below 0.2 kHz. Finally, to maximize the frequency spacing between active and inactive transmons, we park the active transmons above the $\ket{0} \leftrightarrow \ket{3}$ transitions of the fluxoniums. Caution must be taken at this point, since the dynamical flux tuning of the transmons in this parameter regime means that the transmon's $\ket{0}\leftrightarrow\ket{1}$ transition will cross the $\ket{0}\leftrightarrow\ket{3}$ transitions of the fluxoniums. However, we emphasize that since the transmon $\ket{1}$ state and the fluxonium $\ket{3}$ states are both non-computational states, and therefore unoccupied, we do not expect residual errors from this crossing.

In this improved parameter regime, the dependence of the frequency of the conditionally driven transition on the state of spectator qubit is only $\sim 10$ kHz, which is an improvement of two orders of magnitude when compared to the systems inspired by Refs. \cite{mit_cz, sidd_eugene}. As a result, the error averaged over the spectator states shown in Fig. \ref{fig:spec1d} remains below $10^{-5}$. Crucially, in this case we have again numerically calibrated the gate only when the spectator qubits are in the $\ket{0}$ state. As shown in Appendix \ref{app:dmrgx}, the eigenstates remain well localized at the center FTF pair in this parameter regime, and we therefore expect negligible leakage to states outside the center FTF pair. In this parameter regime, the minimum frequency separation $\xi_{ij}$ is only 42 MHz, while for the parameters used in Refs. \cite{mit_cz, sidd_eugene} the minimum frequency separation is 152 MHz and 68 MHz respectively. Naturally, this results in a larger leakage rate. To suppress this leakage, we make use of the hybridization structure of the eigenstates. We apply the gate by driving the $\ket{000}\leftrightarrow\ket{010}$ transition through the charge operator of one of the fluxoniums instead of through the transmon's charge operator. We always drive the fluxonium which has the largest detuning between its $\ket{0}\leftrightarrow\ket{3}$ transition and the transmon's qubit transition which, for the purpose of explaining how the operation works, we assume to be the left fluxonium. That implies we set $\hat{\tilde{D}}=\hat{\tilde{n}}_\text{F1}$ in \mbox{Eq. \eqref{eq:three_node_dh}}. By driving the left fluxonium, the leakage channels $\ket{10i}\leftrightarrow\ket{11i}$ are naturally suppressed, as they do not hybridize with the left fluxonium and therefore do not couple through its charge operator. To suppress the leakage channel $\ket{001}\leftrightarrow\ket{011}$, we will employ the DRAG technique, which is applicable here since we aim to suppress a single nearly-resonant transition frequency in the spectral density of the pulse. We further suppress leakage of $\ket{100}\leftrightarrow\ket{110}$ by adding a small linear compensation drive to the right fluxonium. Full details regarding this drive scheme and its calibration are detailed in \mbox{Appendix \ref{app:calibration}}.

\begin{figure}[b]
\centering
\includegraphics[width=\linewidth]{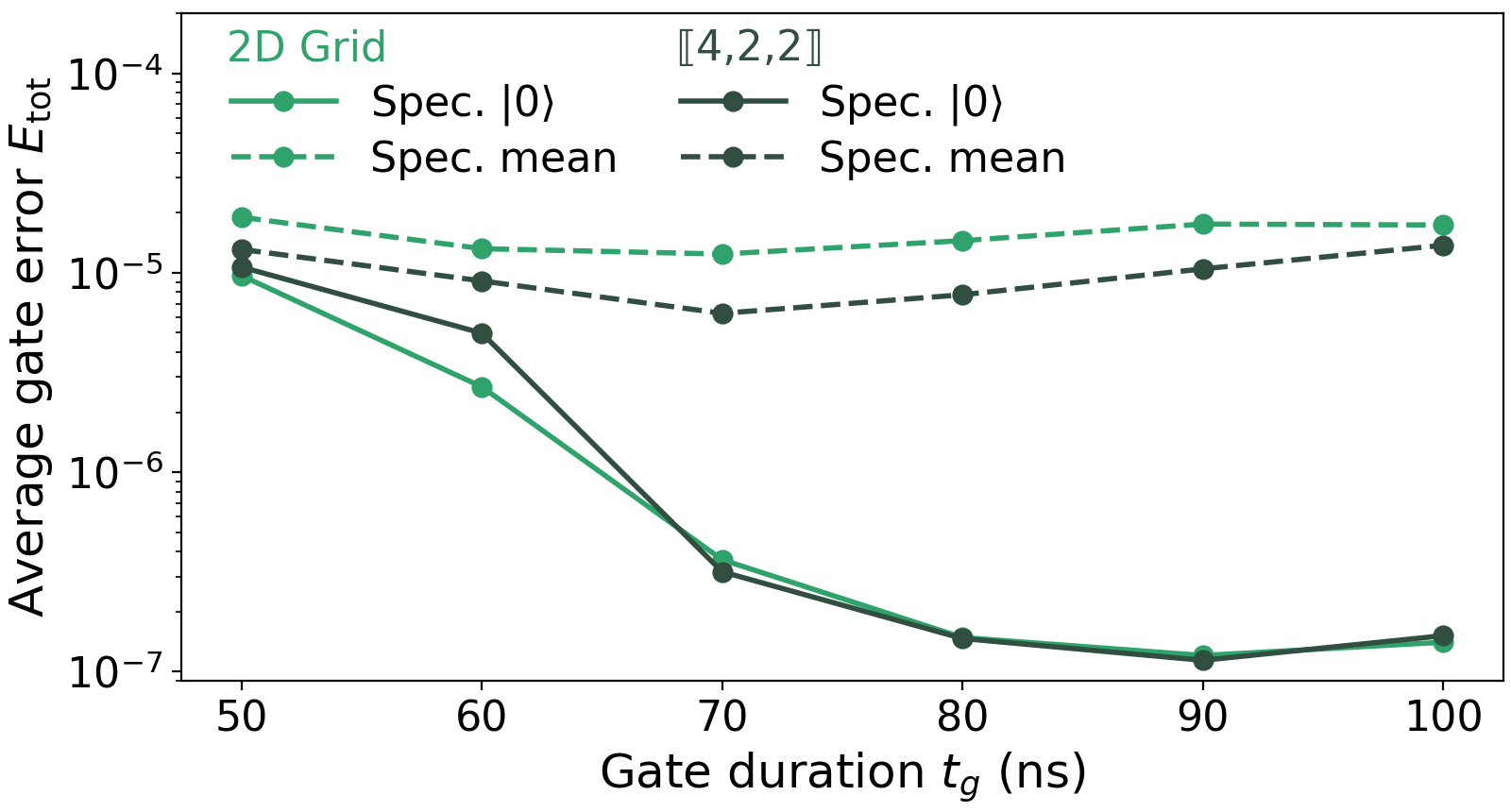}
\caption{\label{fig:spec2d} Spectator errors for the 2D qubit architectures illustrated in Fig. \ref{fig:grid1d}. For each system, the two-qubit operation is calibrated with all spectators in $\ket{0}$ (solid lines), and then the error is averaged over all spectator configurations (dashed lines).}
\end{figure}

Next, we analyze the crosstalk errors in two 2D qubit architectures. We study crosstalk errors from spectator qubits in a 2D grid that could be used as a building block for error correcting surface codes \cite{surfacecode_eth, surfacecode_delft, surfacecode_google2, surfacecode_google}, and a 2D architecture that can be used for a $\llbracket 4,2,2 \rrbracket$ error correcting code with two auxiliary qubits \cite{422_gottesman, 422_takita}. These architectures are shown in \mbox{Figs. \hyperref[fig:grid1d]{\ref*{fig:grid1d}(b)} and \hyperref[fig:grid1d]{\ref*{fig:grid1d}(c)}}, respectively, and the full set of parameters used in this work is available in \mbox{Appendix \ref{app:system_params}}. For both architectures, we also set $g_\text{FT}/2\pi=200$ MHz. As also done for the 1D architecture, we tune inactive transmons down in frequency, and place them around $300$ MHz above the $\ket{1}\leftrightarrow\ket{2}$ transition of the fluxoniums. In the 2D grid, we consider a two-qubit operation performed on the center FTF pair (sites 8, 9 and 10), and for the $\llbracket 4,2,2 \rrbracket$ architecture we consider a two-qubit operation performed on the left-top FTF pair (sites 1, 2 and 3). We calibrate the operation with all spectators in $\ket{0}$, and compute the error averaged over all spectator configurations (i.e. $2^6$ configurations for the 2D grid, and $2^4$ configurations for the $\llbracket 4,2,2 \rrbracket$ architecture). The results of these simulations are shown in \mbox{Fig. \ref{fig:spec2d}}, where we see that the spectator errors remain well below $10^{-4}$ for all gate durations $50$ ns $\leq t_g \leq 100$ ns, thus, providing strong evidence that our proposed parameter regime can serve as a scalable architecture for fluxonium qubits.

\section{Model Imperfections}
\label{sec:model_imperfections}
\begin{figure}[b]
\centering
\includegraphics[width=\linewidth]{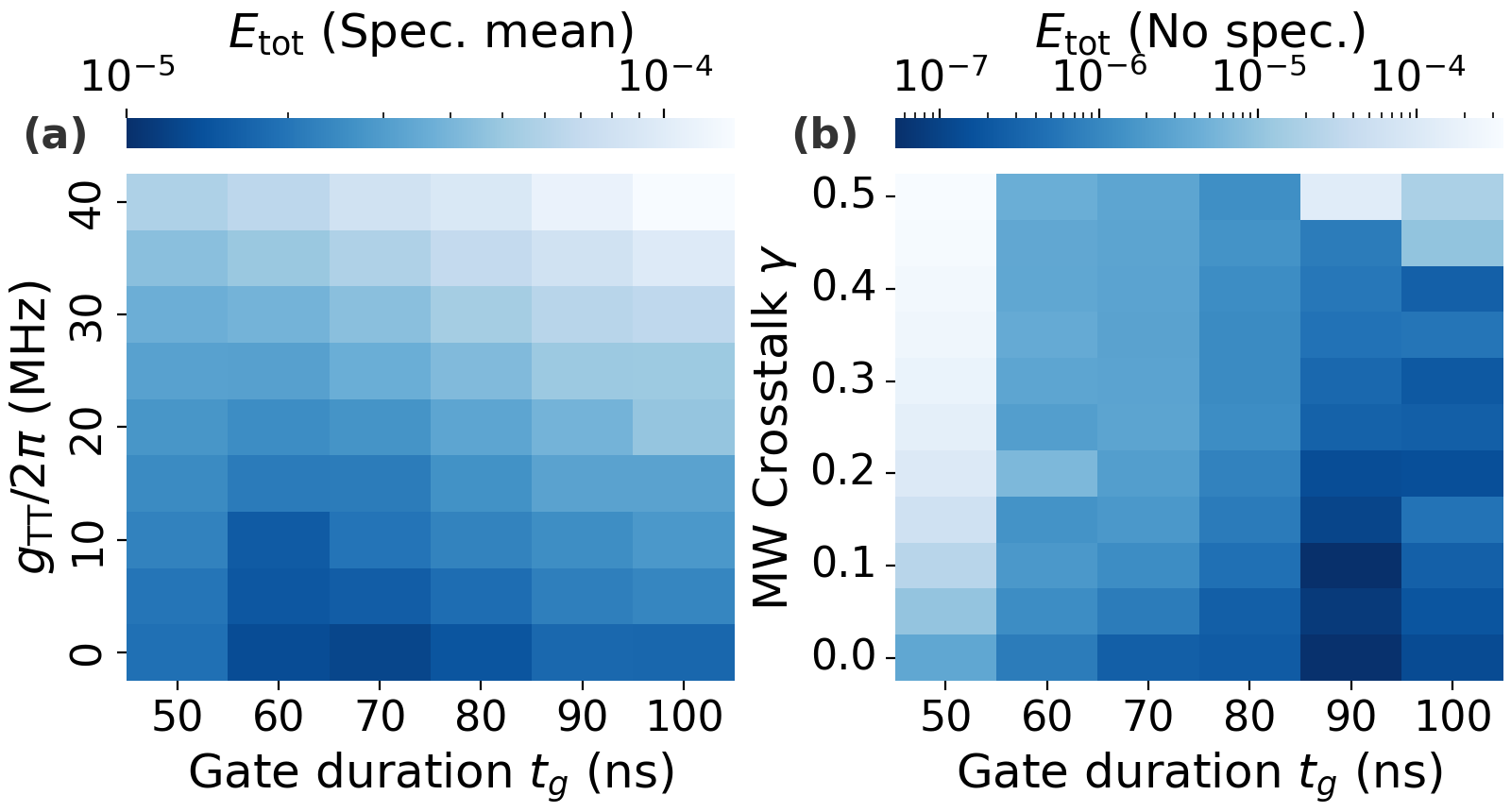}
\caption{\label{fig:model_imperfections} (a) Total error $E_\text{tot}$ averaged over all spectator configurations for the 2D grid system as a function of the gate duration $t_g$ and the capacitive coupling between neighboring transmons $g_\text{TT}$. (b) Total error $E_\text{tot}$ of the bare center FTF pair as a function of the gate duration $t_g$ and the microwave (MW) crosstalk strength $\gamma$.}
\end{figure}

In this section, we analyze the influence of direct capacitive coupling between the transmon couplers and microwave crosstalk on the gate error. First, we add a direct capacitive coupling between all pairs of nearest-neighbor transmons such that the Hamiltonian $\hat{H}_\text{TT}^{ij} = \hbar g_\text{TT} \hat{n}_\text{T$_i$}\hat{n}_\text{T$_j$}$ directly couples transmons $i$ and $j$ with a coupling strength $g_\text{TT}$. For the 2D grid, we see in \mbox{Fig. \hyperref[fig:model_imperfections]{\ref*{fig:model_imperfections}(a)}} that the error, when averaged over all spectator states, remains below $10^{-4}$ even when the transmons are coupled strongly to each other. Here, we have numerically calibrated the gates independently for each $g_\text{TT}$ with all spectator qubits in the $\ket{0}$ state. Noticeably, for $g_\text{TT}/2\pi \geq 20$ MHz the gate error improves for shorter gate durations. This occurs because the derivative of the conditional phase with respect to the drive detuning increases for longer gate durations. As a result, frequency deviations lead to larger conditional phase errors for longer gates.

We recall that in the gate scheme proposed in this work, the operation is applied by driving one of the fluxoniums and a small compensation drive is applied to the other fluxonium. In a realistic device architecture, we expect some capacitive coupling between each fluxonium's drive line and the transmon coupler \cite{mit_cz, sidd_eugene}, which causes effective microwave crosstalk. Here, we simulate the gate error as a function of the microwave crosstalk strength $\gamma$, which is defined as the ratio between the coupling strength between the charge line and the transmon and between the charge line and the fluxonium. We assume $\gamma$ to be the same for both fluxoniums. The exact drive Hamiltonian is detailed in Appendix \ref{app:calibration}.

To isolate the effect of microwave crosstalk from other spectator effects, we compute the total error $E_\text{tot}$ of the bare center FTF pair of the 2D grid system (i.e. no spectator systems are present in these simulations). As shown in Fig. \hyperref[fig:model_imperfections]{\ref*{fig:model_imperfections}(b)}, the influence of microwave crosstalk depends on the gate duration. For $t_g=50$ ns, the gate error increases significantly to $\gtrsim 10^{-4}$, and the dominant error channels in this operation are leakage from $\ket{000} \rightarrow \ket{003}$ and leakage from $\ket{101} \rightarrow \ket{111}$. For the 60 ns, 70 ns and 80 ns gate duration the error increases only slightly with increasing $\gamma$, but remains below $10^{-5}$. For 90 ns and 100 ns the gate error also remains low, but increases when $\gamma \gtrsim 0.4$. We conclude that the gate scheme is moderately resilient against microwave crosstalk, and that the precise effects may depend strongly on the gate duration.

\section{Discussion}
\label{sec:outlook}
In this work, we have investigated spectator errors that arise when scaling-up fluxonium qubit architectures with transmon couplers. We found that spectator errors limit the two-qubit gate fidelity to below 90\% when trivially scaling-up the FTF architecture based on the parameters used in Refs. \cite{mit_cz,sidd_eugene}. By modifying the parameter regime and dynamically tuning inactivate transmons to an off point, we showed that these spectator errors can be suppressed to below $10^{-4}$ in 1D and 2D qubit architectures where each two-qubit operation is influenced by up to six spectator fluxoniums. We further showed that the two-qubit gate operation is robust against direct capacitive coupling between the transmons and moderately resilient against microwave crosstalk. 

\section*{Acknowledgments}
The authors acknowledge support from the Dutch Research Council (NWO). The authors further acknowledge the use of computational resources of the DelftBlue supercomputer, provided by Delft High Performance Computing Centre \cite{DHPC2024}. The authors thank all members of the Andersen lab, Isidora Araya Day, Sebastian Miles, Warre Missiaen and Anton Akhmerov for insightful discussions.

\section*{Data Availability}
All numerical data is available through \cite{data_repository} and the code used for the numerical simulations and data processing is available through \cite{github_repo}.

\appendix

\section{System Parameters}
\label{app:system_params}
\begin{figure}[t]
\centering
\includegraphics[width=\linewidth]{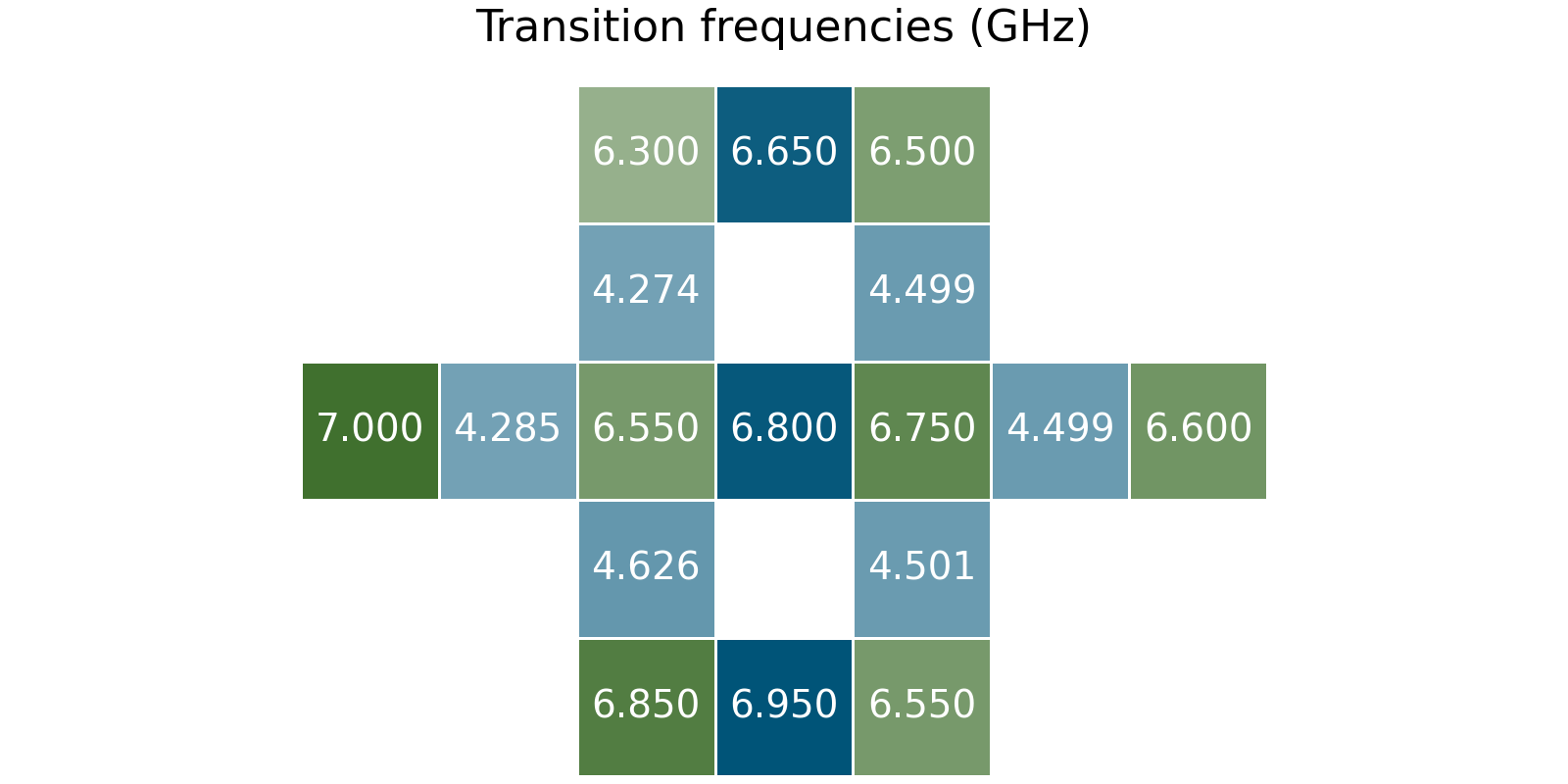}
\caption{\label{fig:eparams} Visualization of the energy parameters used for the 2D grid. For the fluxoniums, the $\ket{0}\leftrightarrow\ket{3}$ transition energy is plotted on a green color scale, and for the transmons the qubit transition frequency is plotted on a blue color scale.}
\end{figure}

The parameters for the 1D system, 2D grid system and $\llbracket 4,2,2 \rrbracket$ system are detailed in Tables \ref{tab:params_1d}, \ref{tab:params_2d} and \ref{tab:params_422} respectively. Tables \ref{tab:params_2d} and \ref{tab:params_422}, in addition to the $E_J$ and $E_{01}$ values used for the DRMG-X simulations, also list the on-parameters $E_J^\text{on}$ and $E_{01}^\text{on}$ for each transmon. The parameters used for the 2D grid system are further shown graphically in Fig. \ref{fig:eparams}. 

The parameters proposed in this work are informed by the following five design rules, which are based on a numerical investigation of the gate scheme for various parameters detailed in Appendix \ref{app:calibration}. We reiterate that we apply the operation by driving the $\ket{000}\leftrightarrow\ket{010}$ transition by driving the fluxonium that has the largest detuning between its $\ket{0}\leftrightarrow\ket{3}$ transition and the transmon's qubit transition. Here, for the purpose of explaining the design rules, we assume that we are driving the operation through the left fluxonium.

\begin{enumerate}
    \item The on position of a transmon should be 50-150 MHz above the $\ket{0}\leftrightarrow\ket{3}$ transition of the right fluxonium to ensure a sufficiently large detuning between the $\ket{000}\leftrightarrow\ket{010}$ and $\ket{001}\leftrightarrow\ket{011}$ transitions. This ensures that the leakage channel $\ket{001}\rightarrow\ket{011}$ can be efficiently suppressed using DRAG.
    \item The on position of a transmon should be at least 250 MHz above the $\ket{0}\leftrightarrow\ket{3}$ transition of the left fluxonium. As this fluxonium is being driven to implement the CZ gate, a too small frequency spacing results in leakage from $\ket{00i}\rightarrow\ket{30i}$. However, the frequency spacing should not be too large to ensure a sufficiently large coupling of the conditionally driven transition $\ket{000}\leftrightarrow\ket{010}$ through the fluxonium's charge operator.
    \item The on position of next-to-nearest-neighbor transmons should be detuned by at least 150 MHz. A smaller detuning may cause resonances between transmons that are at their on position simultaneously. Nearest-neighbor transmons can have the same on frequency, as they are never used simultaneously to apply a two-qubit operation. 
    \item When a transmon is not used to apply a two-qubit operation, it is tuned down in frequency to approximately 300 MHz above the $\ket{1}\leftrightarrow\ket{2}$ transition of the fluxoniums. 
    \item For the individual fluxoniums, we fixed $E_C/h=1$ GHz and picked the $E_J$ and $E_L$ parameters of the fluxoniums to ensure that the frequency gap between the $\ket{1}\leftrightarrow\ket{2}$ and $\ket{0}\leftrightarrow\ket{3}$ is sufficiently large, as this determines the detuning between on and off transmon couplers. 
\end{enumerate}

\section{DMRG-X Implementation}
\label{app:dmrgx}
In this work, we use a two-site density matrix renormalization group for excited states (DMRG-X) \cite{DMRGX} algorithm to compute the eigenvalues and eigenstates of the system. In this algorithm, the state is represented as a matrix-product state (MPS), which is updated iteratively until it converges to an eigenstate of the system. Throughout this iterative optimization, the bond-dimensions of the MPS are increased until the optimization converges in the variance of the Hamiltonian: $\text{Var}(\hat{H},\psi) = \bra{\psi}\hat{H}^2\ket{\psi} - (\bra{\psi}\hat{H}\ket{\psi})^2$. We truncate the optimization at $\text{Var}(\hat{H}) < 10^{-9}$ $\text{GHz}^2$. Numerically, we find that this truncation corresponds to an absolute error in the eigenenergy below 0.1 Hz. Most of the computational complexity of DMRG-X arises from the Lanczos algorithm \cite{lanczos, saad2003}, which is an iterative eigensolver that is used to compute the eigenstates of the effective two-site Hamiltonians. We speed-up the calculation by using CuPy \cite{cupy_learningsys2017} and by running the algorithm on an NVIDIA A100 GPU provided by Ref. \cite{DHPC2024}.

To generate the matrix-product operator that represents the Hamiltonian of the system, we first diagonalize each individual transmon and fluxonium. The transmons are diagonalized in the charge-number basis with a charge-number cutoff of 50 (101 charge states total). The fluxoniums are diagonalized in the harmonic-oscillator basis with 100 levels. After diagonalization, the systems are transformed to the eigenbasis, keeping only the lowest three eigenstates for the transmons and the lowest six eigenstates of the fluxoniums. 

Due to the strong delocalization of the non-computational eigenstates of the systems inspired by \mbox{Refs. \cite{mit_cz, sidd_eugene}}, convergence of DMRG-X (to the correct eigenstate) is not guaranteed. We employ several techniques to ensure and verify that the algorithm converges to the correct eigenstate. To ensure reliable convergence it is imperative to provide DMRG-X with an accurate guess of the eigenstate. In the 1D systems studied in this work, we generate these guesses by computing the eigenvectors of smaller subsystems using exact diagonalization. Specifically, we split the system into five subsystems: the left-most FTF pair, the center FTF pair, the right-most FTF pair, and the two remaining transmons that sit in between these three FTF pairs. The eigenstates of these five subsystems are then inserted into a  single MPS which forms the initial guess for DMRG-X. A similar methodology is employed for the 2D systems studied in this work.

For the parameters corresponding to Refs. \cite{mit_cz, sidd_eugene}, significant delocalization of the non-computational states causes the guessed MPS to not be sufficiently close to the actual eigenstate, and the algorithm tends to converge to a different eigenstate. To resolve this problem, we adiabatically ramp-up the coupling strengths between the five subsystems from 0 to 100\%. This implies that we ramp $g_\text{FT}$ between FX2 and T2, T2 and FX3, FX4 and T4, and T4 and FX5 and $g_\text{FF}$ between FX2 and FX3 and between FX4 and FX5 (for the site labeling, we refer to Table \ref{tab:params_1d}). When the algorithm detects that it does not converge or has converged to an incorrect state, the step-size of the ramp is decreased. To improve the convergence rate of the algorithm, it is important to detect that the algorithm will not converge (to the correct state) as early as possible. We ensure this by (1) Setting an upper bound on the variance of the guessed state and \mbox{(2) Setting} and upper bound on the allowed change in the eigenstate of the local two-site effective Hamiltonians. 

\begin{figure}[t]
\centering
\includegraphics[width=\linewidth]{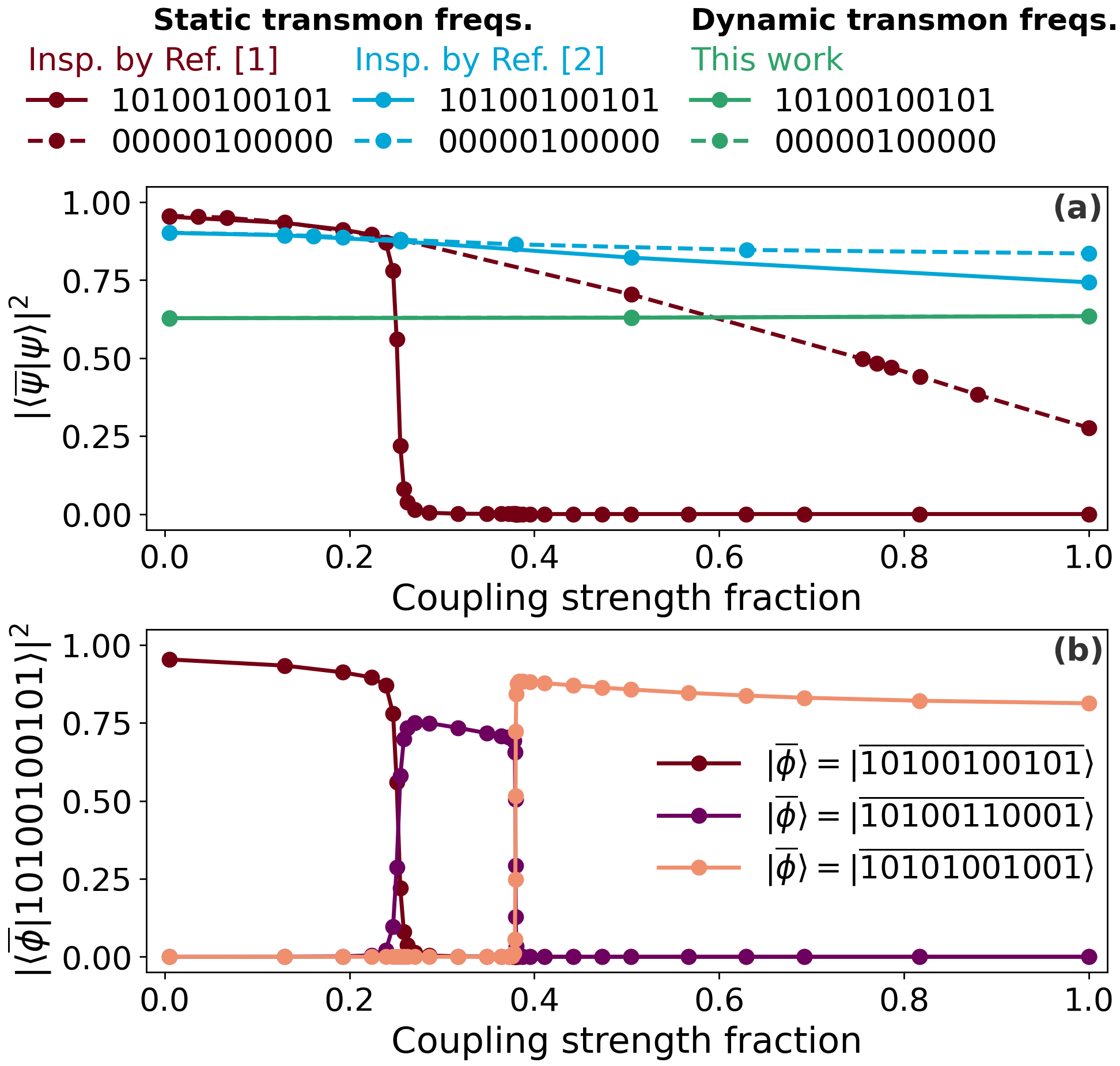}
\caption{\label{fig:adiabatic_ramp} (a) Overlap between the computed eigenstate $\ket{\psi}$ and the bare eigenstate $\ket{\overline{\psi}}$ as a function of the coupling strength fraction for two different states. (b) Hybridization of the $\ket{10100100101}$ state for the parameters based on \mbox{Ref. \cite{mit_cz}} as a function of the coupling strength fraction. As a result of resonances, the state hybridizes very strongly with two other bare eigenstates.}
\end{figure}

Throughout the optimization we monitor the overlap between the optimized MPS and the initial guess and between the optimized MPS and the bare eigenstate of the system. In Fig. \hyperref[fig:adiabatic_ramp]{\ref*{fig:adiabatic_ramp}(a)} we plot the overlap between the optimized eigenstate $\ket{\psi}$ and the bare eigenstate $\ket{\overline{\psi}}$ as a function of the coupling strength fraction between the five subsystems. Here, the bare eigenstate $\ket{\overline{\psi}}$ is an eigenstate of the system without any coupling terms. The notation used for the kets $\ket{\cdot}$ follows the site ordering for the 1D system shown in \mbox{Fig. \hyperref[fig:grid1d]{\ref*{fig:grid1d}(a)}}. We plot the overlap for two states that both have the center FTF pair in $\ket{010}$, but have all spectator fluxoniums either in $\ket{0}$ or in $\ket{1}$. For the parameters based on Refs. \cite{mit_cz,sidd_eugene}, the hybridization of the states changes significantly when the coupling strength increases. Furthermore, the hybridization depends on the state of the spectator qubits, indicating that the matrix element of the state with the drive operator will also depend on the state of spectator qubits, which results in spectator errors. For the parameters proposed in this work, we see that the eigenstate does not depend on the coupling strength nor on the state of the spectator qubits, indicating that it does not delocalize strongly outside the center FTF pair. 

Figure \hyperref[fig:adiabatic_ramp]{\ref*{fig:adiabatic_ramp}(a)} shows that the $\ket{10100100101}$ state for the parameters inspired by \mbox{Ref. \cite{mit_cz}} completely loses character and has a negligible overlap with its bare state after a certain coupling strength. This is most likely due to resonances that occur in the system, and further emphasizes the need for the adiabatic ramp in this implementation of DMRG-X. In Fig. \hyperref[fig:adiabatic_ramp]{\ref*{fig:adiabatic_ramp}(b)} we plot the overlap of this state with two other bare states, and see that it hybridizes significantly with the $\ket{\overline{10100110001}}$ and $\ket{\overline{10101001001}}$ states.

\begin{figure}[t]
\centering
\includegraphics[width=\linewidth]{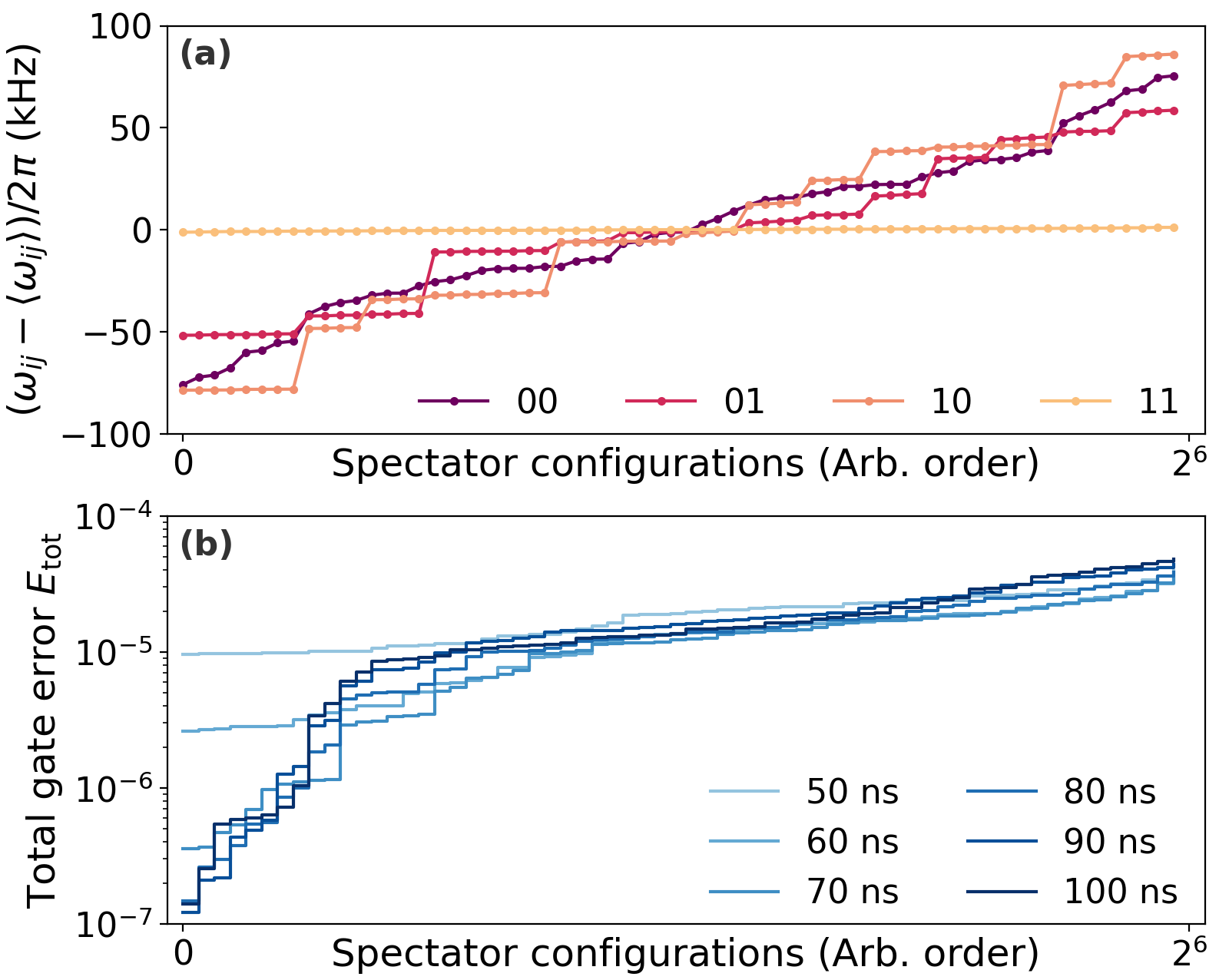}
\caption{\label{fig:app_dmrg} (a) Deviations of the transmon's qubit transition frequency $\omega_{ij}$ as a function of the state of the fluxonium qubits $ij$ for all spectator configurations for the 2D grid system. \mbox{(b) Total} gate error $E_\text{tot}$ for all gate durations and for all spectator configurations. In panels (a) and (b) the order of the spectator configurations is arbitrary (Arb.) as each trace is individually sorted on the $y$ axis.}
\end{figure}

For the 2D grid system, we computed the eigenvalues and eigenstates relevant for the two-qubit gate operation for all $2^6$ spectator configurations. In Fig. \hyperref[fig:app_dmrg]{\ref*{fig:app_dmrg}(a)} we plot the transmon's transition frequency $\omega_{ij} = (E_{i1j}-E_{i0j})/\hbar$ for all spectator configurations. For the states that have at least one fluxonium in the ground state, the transmon's qubit frequency deviates by at most 100 kHz from its frequency averaged over all spectator configurations $\langle \omega_{ij} \rangle$. When the fluxoniums are in $ij=11$, the frequency deviations remain approximately below 1 kHz. In \mbox{Fig. \hyperref[fig:app_dmrg]{\ref*{fig:app_dmrg}(b)}} we plot the gate error for all spectator configurations after calibration the operation with all spectators in $\ket{0}$. For all gate durations and spectator configurations, the error remains below $10^{-4}$.

\section{Gate Calibration}
\label{app:calibration}
To suppress crosstalk errors, we reduced the coupling strength $g_\text{FT}$ and parked the transmon above the $\ket{0}\leftrightarrow\ket{3}$ transition of the fluxoniums. This choice of parameter regime results in a smaller minimum frequency separation $\xi_{ij}$, which increases the leakage rate of the operation. While the minimum frequency separation of the systems in Refs. \cite{mit_cz, sidd_eugene} are 152 MHz and 68 MHz respectively, the minimum frequency separation for the parameter regime proposed in this work is only 42 MHz. As a result, the operation is limited by three leakage channels: $\ket{i0j}\leftrightarrow\ket{i1j}$ for $ij$ not equal to the conditionally driven transition. It is unclear how these leakage channels can be suppressed simultaneously with conventional pulse shaping techniques such as DRAG \cite{drag1,drag2,hd-drag}, as they are often designed to correct for only one leakage channel.

In this work, we leverage the delocalization of the transmon's wavefunction to suppress two of the three leakage channels. Since the transmon's frequency is close in energy to the $\ket{0}\leftrightarrow\ket{3}$ transitions of the fluxoniums, but far detuned from the fluxoniums' $\ket{1}\leftrightarrow\ket{2}$ transitions, its wavefunction will delocalize to a fluxonium when the fluxonium is in the $\ket{0}$ state, and not when it is in the $\ket{1}$ state. Therefore, if we implement the two-qubit operation by driving the $\ket{000}\leftrightarrow\ket{010}$ through the charge operator of the left fluxonium, the leakage transitions $\ket{100}\leftrightarrow\ket{110}$ and $\ket{101}\leftrightarrow\ket{111}$ are immediately suppressed as these transitions do not couple through the charge operator of the left fluxonium. This leaves $\ket{001}\leftrightarrow\ket{011}$ as the only leakage transition, which can be suppressed using DRAG. A numerical investigation of this gate scheme revealed that small leakage from $\ket{100}\rightarrow\ket{110}$ can still remain, most likely as a result of the direct capacitive coupling between the fluxoniums. We further suppress this leakage channel by adding a small compensation drive to the right fluxonium. 

The operation could be applied by driving any of the two fluxoniums. However, the leakage is maximally suppressed by driving the fluxonium that has the largest detuning between the $\ket{0}\leftrightarrow\ket{3}$ transition and the transmon's $\ket{0}\leftrightarrow\ket{1}$ transition, as this maximizes the detuning between the $\ket{000}\leftrightarrow\ket{010}$ transition and the leakage transition that is suppressed using DRAG. By driving the fluxonium instead of the transmon we introduce two more leakage channels into the operation, namely $\ket{000}\leftrightarrow\ket{300}$ and $\ket{001}\leftrightarrow\ket{301}$. We ensure these leakage transitions are suppressed by designing them to be sufficiently far detuned in frequency from the driven transition.

The full drive Hamiltonian of this system previously defined in Eq. \eqref{eq:three_node_dh} is given by

\begin{equation}
\label{eq:drive_hamiltonian}
    \hat{\tilde{H}}_D/\hbar = \Omega \big(\mathcal{E}_I(t)\cos(\omega_dt) + \lambda\mathcal{E}_Q(t)\sin(\omega_dt) \big)\hat{\tilde{D}}
\end{equation}

\noindent Here, $\Omega$ is the drive strength, $\lambda$ the DRAG parameter, $\omega_d$ the drive frequency, $\hat{\tilde{D}}$ the drive operator, $\mathcal{E}_I(t)$ the in-phase pulse envelope previously defined in Eq. \eqref{eq:consine_pe} and $\mathcal{E}_Q(t)$ the quadrature pulse envelope defined as

\begin{equation}
\label{eq:pulse_envelopes}
\begin{split}
    & \mathcal{E}_Q(t) = \frac{\partial \mathcal{E}_I(t)}{\partial t} = \frac{\pi}{t_g}\sin\left(\frac{2\pi}{t_g}t\right).
\end{split}
\end{equation}

\noindent We reiterate that the $\tilde{\cdot}$ indicates that the operator has been transformed to the eigenbasis. When the left fluxonium is driven, the drive operator $\hat{\tilde{D}}$ is equal to

\begin{equation}
    \hat{\tilde{D}} \mapsto \hat{\tilde{n}}_\text{F$_1$} + \eta\hat{\tilde{n}}_\text{F$_2$} + \gamma(1 + \eta)\hat{\tilde{n}}_\text{T}.
\end{equation}

\noindent Here, $\eta$ is the relative drive strength used to suppress the $\ket{100}\leftrightarrow\ket{110}$ transition, and $\gamma$ is the microwave crosstalk strength used in Sec. \ref{sec:model_imperfections}.

The following procedure is used to calibrate the gate operation assuming that the left fluxonium is being driven:

\begin{enumerate}
    \item Calibration of the drive frequency and drive strength. For a range of detunings, the drive strength is optimized to minimize the leakage $1-\vert\bra{000}U\ket{000}\vert^2$. For each detuning, the conditional phase is computed, and the results are fitted with a linear function to find the approximate detuning that provides a conditional phase of $\pi$. The procedure is repeated until the conditional phase error converges to below $10^{-7}$.
    \item The DRAG parameter $\lambda$ is optimized to minimize leakage $1-\vert\bra{001}U\ket{001}\vert^2$.
    \item The relative drive strength $\eta$ is optimized to minimize the leakage $1-\vert\bra{100}U\ket{100}\vert^2$.
    \item The drive frequency and drive strength are optimized with the same procedure as step 1.
\end{enumerate}

\noindent Step 2 is skipped if the leakage from $\ket{001}$ is already below $10^{-7}$. Similarly, step 3 is skipped if the leakage from $\ket{100}$ is below $10^{-7}$. Steps 2-4 are repeated at most two times. Calibration for a system in which the right fluxonium is driven is, of course, analogous, with the notable change that the DRAG parameter is used to minimize leakage from the $\ket{100}$ state and the relative drive strength to minimize leakage from the $\ket{001}$ state. 

\begin{figure}[t]
\centering
\includegraphics[width=\linewidth]{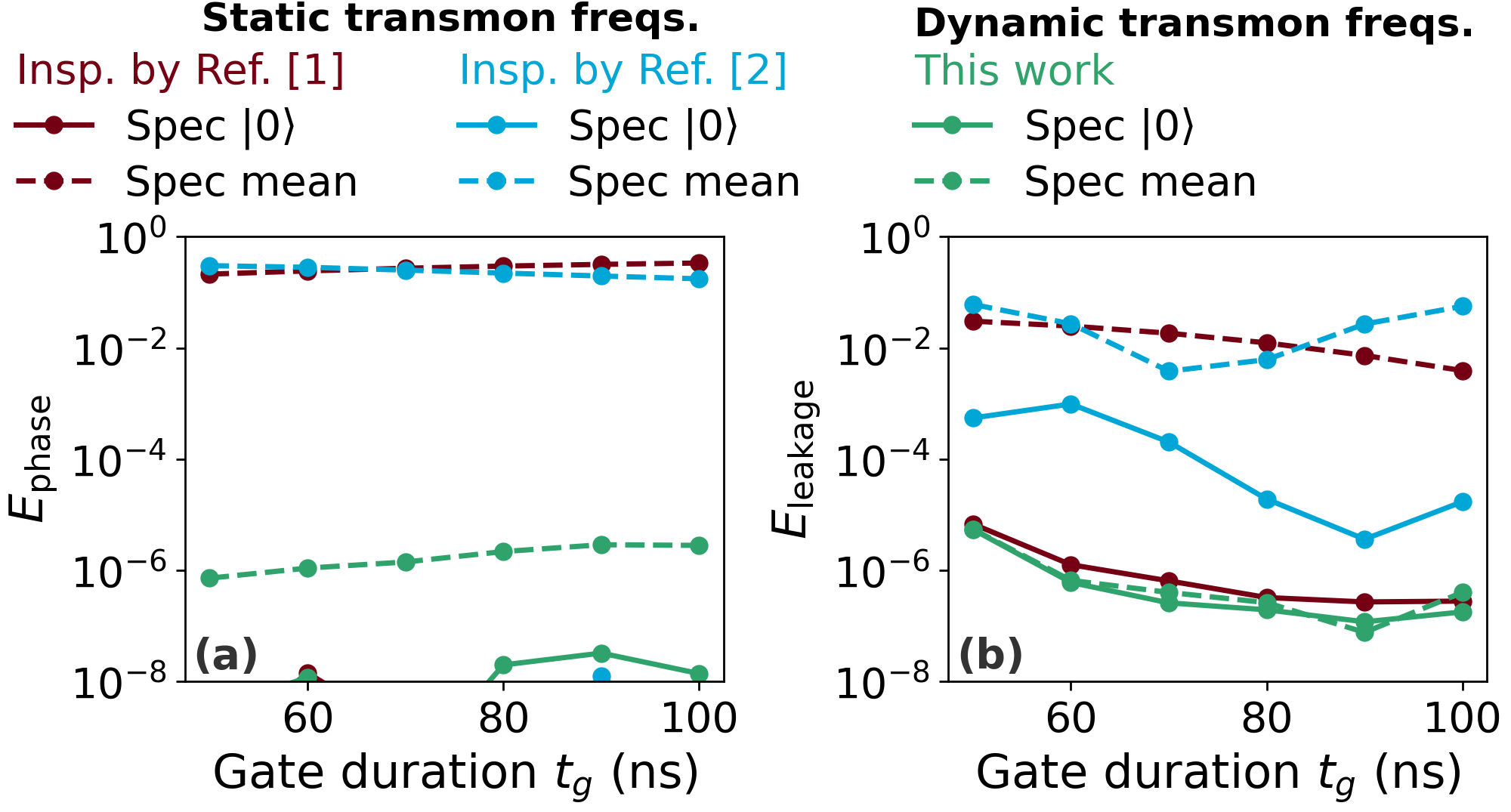}
\caption{\label{fig:spec1d_indv} (a), (b) Individual contributions from the phase error and leakage error, respectively, to the total error plotted in \mbox{Fig. \ref{fig:spec1d}}. The phase errors for the calibrated operations with all spectators in $\ket{0}$ are $\lesssim 10^{-8}$ and are therefore not fully visible in panel (a).}
\end{figure}

We emphasize that we only employ this calibration scheme for the systems with the parameters proposed in this work. For the time evolutions performed on the systems with the parameters inspired by Refs. \cite{mit_cz, sidd_eugene} we only calibrate a simple cosine pulse envelope and drive the $\ket{101}\leftrightarrow\ket{111}$ transition directly though the transmon's charge operator, i.e. $\lambda=\eta=0$ and $\hat{\tilde{D}}=\hat{\tilde{n}}_\text{T}$. The pulse is subsequently calibrated by optimizing the drive strength to minimize the leakage $1 - \vert\bra{101}U\ket{101}\vert^2$ and optimizing the drive frequency to ensure the conditional phase is $\pi$ with a similar procedure as described in step 1 of the calibration protocol detailed above. For the 1D architecture, the individual contributions from coherent phase errors and leakage errors to the total gate error are shown in Fig. \ref{fig:spec1d_indv}.

\begin{figure}[t]
\centering
\includegraphics[width=\linewidth]{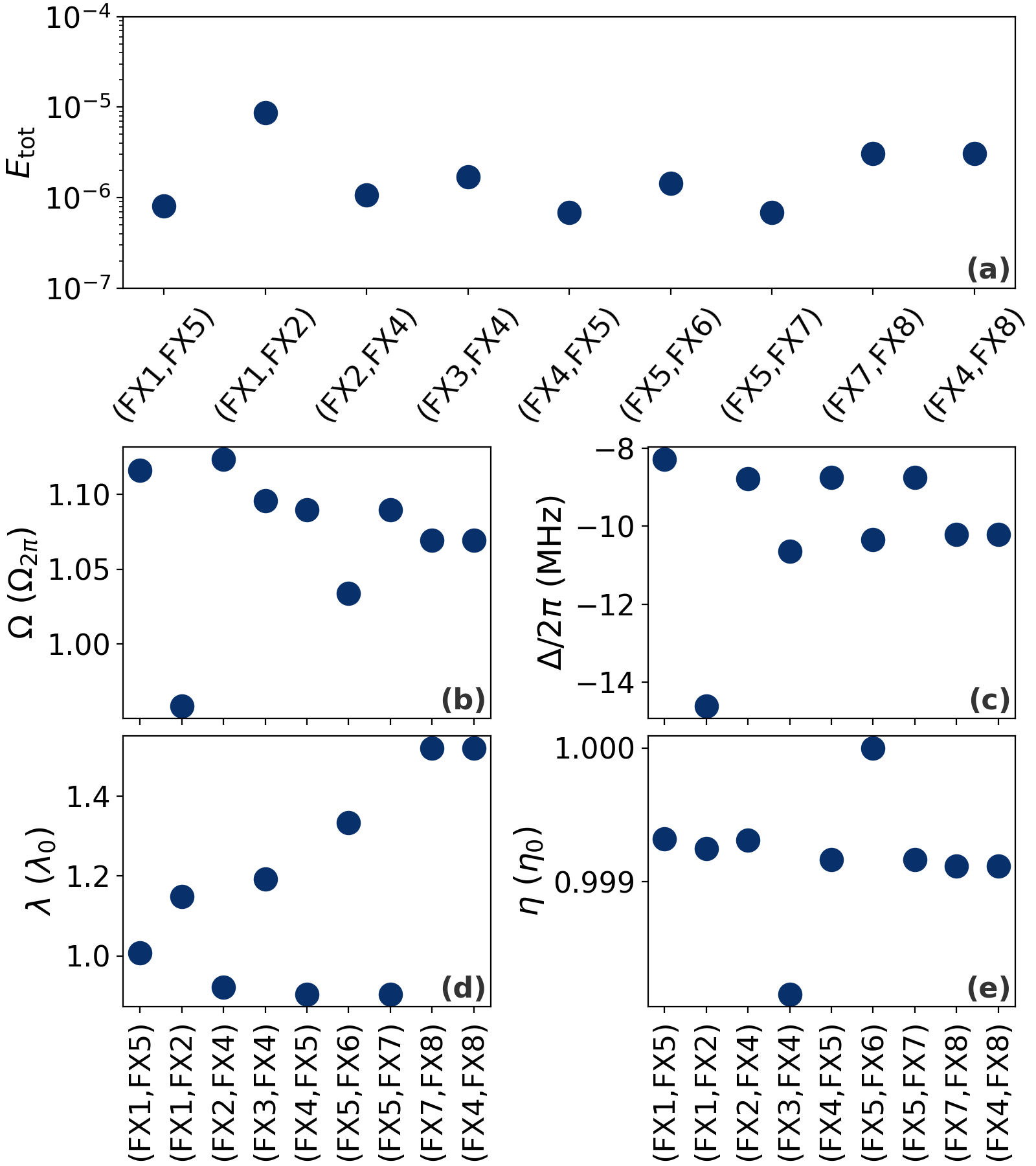}
\caption{\label{fig:all_pairs} (a) Total error $E_\text{tot}$ for all bare FTF pairs in the 2D grid system. (b)-(e) Calibrated drive parameters. In panels (b), (d) and (e) the drive parameters are plotted in units of $\Omega_{2\pi}$, $\lambda_0$ and $\eta_0$ respectively which are defined in the main text.}
\end{figure}

To demonstrate the versatility of the gate scheme to the varying parameters of each FTF pair, we optimize a 60 ns CZ gate for each bare FTF pair in the 2D grid system of which the results are shown in Fig. \ref{fig:all_pairs}, where we see that the gate error is below $10^{-5}$ for each FTF pair. We further plot the optimized pulse parameters in Figs. \hyperref[fig:all_pairs]{\ref*{fig:all_pairs}(b)-(e)}. We plot the optimized drive strength $\Omega$ in units of the drive strength $\Omega_{2\pi}$, which is the drive strength that would perform a $2\pi$ rotation of the same duration on the qubit transition of the driven fluxonium (if it were a perfect, uncoupled two-level system). The DRAG parameter $\lambda$ and relative drive strength are plotted in units of their theoretically predicted values, which, assuming the left fluxonium is being driven, are given by

\begin{equation}
\begin{split}
    &\lambda_0 = \frac{1}{\omega_d - (\omega_{011}-\omega_{001})}, \\
    &\eta_0 = -\frac{\bra{110}\hat{n}_{\text{F}_1}\ket{100}}{\bra{110}\hat{n}_{\text{F}_2}\ket{100}}.
\end{split}
\end{equation}

\FloatBarrier

\bibliography{apssamp}

\begin{table*}[t]
\caption{\label{tab:params_1d} Parameters used for the 1D system consisting of six fluxoniums and five transmons. For Ref. \cite{mit_cz}, the parameters of sites 5, 6 and 7 are copied from Device B at the flux-point $\varphi_\text{ext}=0$. For \mbox{Ref. \cite{sidd_eugene}}, the parameters of sites 5, 6 and 7 are copied from the flux-point $\varphi_\text{ext}=0$. For Ref. \cite{mit_cz} the coupling strengths are $g_\text{FT}/2\pi=550$ MHz and $g_\text{FF}/2\pi=120$ MHz, for \mbox{Ref. \cite{sidd_eugene}} $g_\text{FT}/h=236$ MHz and $g_\text{FF}/2\pi=40$ MHz and in this work we use $g_\text{FT}/2\pi=200$ MHz and $g_\text{FF}/2\pi=20$ MHz. All parameters are in units of GHz.}
\begin{ruledtabular}
\begin{tabular}{l|ccccccccccc}
    Site & 1 & 2 & 3 & 4 & 5 & 6 & 7 & 8 & 9 & 10 & 11 \\
    System & FX1 & T1 & FX2 & T2 & FX3 & T3 & FX4 & T4 & FX5 & T5 & FX6 \\
    \hline
    Ref. \cite{mit_cz} & & & & & & & & & & \\
    $E_C/h$ & 1.41 & 0.30 & 1.33 & 0.30 & 1.41 & 0.30 & 1.33 & 0.30 & 1.41 & 0.30 & 1.33 \\
    $E_J/h$ & 5.85 & 16.4 & 5.55 & 16.2 & 5.7 & 16.0 & 5.4 & 15.8 & 5.55 & 15.6 & 5.25 \\
    $E_L/h$ & 0.88 & & 0.60 & & 0.88 & & 0.60 & & 0.88 & & 0.66 \\
    $E_{01}/h$ & 0.397 & 5.957 & 0.256 & 5.919 & 0.422 & 5.880 & 0.273 & 5.841 & 0.449 & 5.802 & 0.292 \\
    $E_{12}/h$ & 5.233 & & 5.166 & & 5.124 & & 5.049 & & 5.016 & & 4.932 \\
    $E_{03}/h$ & 8.354 & & 7.525 & & 8.320 & & 7.474 & & 8.287 & & 7.425 \\
    \hline
    Ref. \cite{sidd_eugene} & & & & & & & & & & & \\
    $E_C/h$ & 0.88 & 0.186 & 0.88 & 0.186 & 0.88 & 0.186 & 0.88 & 0.186 & 0.88 & 0.186 & 0.88 \\
    $E_J/h$ & 5.170 & 21.5 & 5.015 & 19.0 & 4.993 & 16.87 & 4.335 & 16.0 & 4.000 & 15.0 & 3.550 \\
    $E_L/h$ & 0.95 & & 0.75 & & 0.50 & & 0.50 & & 0.40 & & 0.35 \\
    $E_{01}/h$ & 0.213 & 5.463 & 0.160 & 5.124 & 0.095 & 4.817 & 0.142 & 4.685 & 0.138 & 4.530 & 0.162 \\
    $E_{12}/h$ & 4.016 & & 4.102 & & 4.346 & & 3.835 & & 3.672 & & 3.365 \\
    $E_{03}/h$ & 6.062 & & 5.827 & & 5.587 & & 5.347 & & 5.101 & & 4.860 \\
    \hline
    This work & & & & & & & & & & & \\
    $E_C/h$ & 1 & 0.25 & 1 & 0.25 & 1 & 0.25 & 1 & 0.25 & 1 & 0.25 & 1 \\
    $E_J/h$ & 4.481 & 23.874 & 4.118 & 10.308 & 4.889 & 24.920 & 5.330 & 11.360 & 4.889 & 25.989 & 5.572 \\
    $E_L/h$ & 0.95 & & 0.85 & & 0.95 & & 1.05 & & 0.95 & & 1.1 \\
    $E_{01}/h$ & 0.416 & 6.650 & 0.432 & 4.274 & 0.333 & 6.800 & 0.308 & 4.501 & 0.333 & 6.950 & 0.293 \\
    $E_{12}/h$ & 3.701 & & 3.529 & & 3.974 & & 4.199 & & 3.974 & & 4.326 \\
    
    $E_{03}/h$ & 6.500 & & 6.300 & & 6.550 & & 6.750 & & 6.550 & & 6.850
\end{tabular}
\end{ruledtabular}
\end{table*}

\begin{table*}[t]
\caption{\label{tab:params_2d} Parameters used for the 2D grid system. For this system, we also use $g_\text{FT}/2\pi=200$ MHz and $g_\text{FF}/2\pi=20$ MHz. All parameters are in units of GHz.}
\begin{ruledtabular}
\begin{tabular}{l|ccccccccccccccccc}
    Site & 1 & 2 & 3 & 4 & 5 & 6 & 7 & 8 & 9 & 10 & 11 & 12 & 13 & 14 & 15 & 16 & 17 \\
    System & T1 & FX1 & T2 & FX2 & T3 & FX3 & T4 & FX4 & T5 & FX5 & T6 & FX6 & T7 & FX7 & T8 & FX8 & T9 \\
    \hline
    $E_C/h$ & 0.25 & 1 & 0.25 & 1 & 0.25 & 1 & 0.25 & 1 & 0.25 & 1 & 0.25 & 1 & 0.25 & 1 & 0.25 & 1 & 0.25 \\
    $E_J/h$ & 11.350 & 4.481 & 23.874 & 4.118 & 10.308 & 5.236 & 10.357 & 4.889 & 24.920 & 5.330 & 11.350 & 5.176 & 11.360 & 4.889 & 25.989 & 5.572 & 11.961 \\
    $E_L/h$ &  & 0.95 &  & 0.85 &  & 1.25 &  & 0.95 &  & 1.05 &  & 0.95 &  & 0.95 &  & 1.1 &  \\
    $E_{01}/h$ & 4.499 & 0.416 & 6.650 & 0.432 & 4.274 & 0.435 & 4.285 & 0.333 & 6.800 & 0.308 & 4.499 & 0.285 & 4.501 & 0.333 & 6.950 & 0.293 & 4.626 \\
    $E_{12}/h$ &  & 3.701 &  & 3.529 &  & 3.985 &  & 3.974 &  & 4.199 &  & 4.176 &  & 3.974 &  & 4.326 &  \\
    $E_{03}/h$ &  & 6.500 &  & 6.300 &  & 7.000 &  & 6.550 &  & 6.750 &  & 6.600 &  & 6.550 &  & 6.850 &  \\
    $E_J^\text{on}/h$ & 24.920 &  & 23.874 &  & 23.531 &  & 27.080 &  & 24.920 &  & 24.920 &  & 24.920 & & 25.989 &  & 25.989 \\ 
    $E_{01}^\text{on}/h$ & 6.800 &  & 6.650 &  & 6.600 &  & 7.100 &  & 6.800 &  & 6.800 &  & 6.800 & & 6.950 &  & 6.950 \\
\end{tabular}
\end{ruledtabular}
\end{table*}

\begin{table*}[t]
\caption{\label{tab:params_422} Parameters used for the $\llbracket 4,2,2 \rrbracket$ system. For this system, we also use $g_\text{FT}/2\pi=200$ MHz and $g_\text{FF}/2\pi=20$ MHz. All parameters are in units of GHz.}
\begin{ruledtabular}
\begin{tabular}{l|ccccccccccccccccc}
    Site & 1 & 2 & 3 & 4 & 5 & 6 & 7 & 8 & 9 & 10 & 11 & 12 & 13 & 14 \\
    System & FX1 & T1 & FX2 & T2 & FX3 & T3 & FX4 & T4 & T5 & FX5 & T6 & T7 & FX6 & T8 \\
    \hline
    $E_C/h$ & 1 & 0.25 & 1 & 0.25 & 1 & 0.25 & 1 & 0.25 & 0.25 & 1 & 0.25 & 0.25 & 1 & 0.25 \\
    $E_J/h$ & 5.330 & 24.920 & 4.889 & 11.960 & 5.572 & 25.989 & 4.889 & 11.349 & 11.359 & 5.176 & 11.970 & 11.980 & 4.481 & 11.369 \\
    $E_L/h$ & 1.05 &  & 0.95 &  & 1.1 &  & 0.95 &  &  & 0.95 &  &  & 0.95 & \\
    $E_{01}/h$ & 0.308 & 6.800 & 0.333 & 4.626 & 0.293 & 6.950 & 0.333 & 4.499 & 4.501 & 0.285 & 4.628 & 4.630 & 0.416 & 4.503 \\
    $E_{12}/h$ & 4.199 &  & 3.974 &  & 4.326 &  & 3.974 &  &  & 4.176 &  &  & 3.701 & \\
    $E_{03}/h$ & 6.750 &  & 6.550 &  & 6.850 &  & 6.550 &  &  & 6.600 &  &  & 6.500 & \\
    $E_J^\text{on}/h$ &  & 24.920 &  & 25.989 &  & 25.989 &  & 24.920 & 24.920 &  & 25.989 & 25.989 &  & 24.920\\
    $E_{01}^\text{on}/h$ &  & 6.800 &  & 6.950 &  & 6.950 &  & 6.800 & 6.800 &  & 6.950 & 6.950 &  & 6.800 \\
\end{tabular}
\end{ruledtabular}
\end{table*}

\end{document}